\documentclass{aa}
\usepackage[varg]{txfonts}
\usepackage{graphicx}
\usepackage{natbib}
\usepackage[backref,breaklinks,colorlinks,citecolor=blue]{hyperref}
\usepackage{booktabs}
\usepackage[para,online,flushleft]{threeparttable}

\begin{document}

\title{Ground-based optical transmission spectrum of the hot Jupiter HAT-P-1\lowercase{b}}

\author{Kamen O. Todorov\inst{\ref{inst1}}
\and Jean-Michel D\'esert\inst{\ref{inst1}}
\and Catherine M. Huitson\inst{\ref{inst2}}
\and Jacob L. Bean\inst{\ref{inst3}}
\and Vatsal Panwar\inst{\ref{inst1}}
\and Filipe de Matos\inst{\ref{inst1}}
\and Kevin B. Stevenson\inst{\ref{inst5}}
\and Jonathan J. Fortney\inst{\ref{inst4}}
\and Marcel Bergmann\inst{\ref{inst6}}
}

\institute{Anton Pannekoek Institute for Astronomy,
University of Amsterdam, Science Park 904, XH 1098 Amsterdam, The Netherlands
\email{ktodorov@uva.nl}\label{inst1}
\and CASA, University of Colorado, 389 UCB, Boulder, CO, 80309-0389, USA
\label{inst2}
\and Department of Astronomy and Astrophysics, University of Chicago, Chicago, IL 60637, USA\label{inst3}
\and Department of Astronomy and Astrophysics, University of California, Santa Cruz, CA 95064, USA\label{inst4}
\and Space Telescope Science Institute, 3700 San Martin Drive, Baltimore, MD 21218, USA\label{inst5}
\and NOAO \ Gemini Observatory, present address Palo Alto, CA, USA\label{inst6}
}

   \date{Received \today; accepted 17 Sep 2019}
\abstract{Time-series spectrophotometric studies of exoplanets during transit
using ground-based facilities are a promising approach to 
characterize their atmospheric compositions. }
{We aim to investigate the transit spectrum of the hot Jupiter HAT-P-1b. We compare
our results to those obtained at similar wavelengths by previous space-based observations. }
{We observed two transits of HAT-P-1b with the 
Gemini Multi-Object Spectrograph (GMOS) instrument on the Gemini North 
telescope using 
two instrument modes covering the 320 -- 800\,nm and 520 -- 950\,nm wavelength ranges. We
used time-series spectrophotometry to construct transit light curves in individual 
wavelength bins and measure the transit depths in each bin. We accounted 
for systematic effects. We addressed potential photometric variability due to 
magnetic spots in the planet's host star with long-term photometric monitoring. 
}
{We find that the resulting transit spectrum is consistent with previous 
Hubble Space Telescope (HST) observations. 
We compare our observations to transit spectroscopy models that marginally favor a clear atmosphere.
However, the observations are also consistent with a flat spectrum, indicating high-altitude clouds.
We do not detect the Na resonance absorption line (589\,nm), and 
our observations do not have sufficient precision to study the resonance line of K at 770\,nm.
}
{We show that even a single Gemini/GMOS transit can provide constraining power
on the properties of the atmosphere of HAT-P-1b to a level comparable to that of HST 
transit studies in the optical when the observing conditions and target and reference star combination are suitable. 
Our 520 -- 950\,nm observations
reach a precision comparable to that of HST transit spectra in a similar wavelength range of 
the same hot Jupiter, HAT-P-1b. However, our GMOS transit between 320 -- 800\,nm suffers from 
strong systematic effects and yields larger uncertainties.
}


\titlerunning{GMOS spectrum of HAT-P-1b}
\authorrunning{Todorov et al.}
\maketitle
%________________________________________________________________

\section{Introduction}
The atmospheres of transiting hot 
exoplanets have been probed through time-series transit
spectroscopy, for instance, using data from the Hubble Space Telescope 
\citep[HST, e.g.,][]{cha02, vid03, dem13, sin16, arc18}. 
From the ground, transit studies have used multi-object spectrographs 
\citep[MOS; e.g.,][]{bea10, bea11, gib13b, ste14, nik16, hui17, rac17, bix19, esp19} and
long-slit spectrographs in low-resolution mode 
\citep[e.g.,][]{sin12, mur14, nor16, mur19}. 
Studies using ground-based high-resolution 
spectrographs \citep[e.g.,][]{red08, sne08, nor18}
have also yielded information on the planetary atmospheres.

We here focus on the MOS technique, 
which relies on time-series spectrophotometry of the target
star-planet system during transit. 
The objective, as with all time-series transit spectroscopy, is to
measure the depth of a transit as a function of wavelength to extract 
a transit spectrum, which carries information about the contents of the 
planetary atmosphere. A reference star is typically observed in another
MOS slit to help mitigate any spectrophotometric systematic effects.
Long-slit spectrographs have also been used. 

This work, along with the study by \citet{hui17}, 
is part of a transit spectroscopy survey using the Gemini/GMOS instrument
that observed nine transiting exoplanets during a total of about 40 transits 
with high spectrophotometric precision in visible wavelengths. 
The results from \citet{hui17} have been used by \citet{bou19} 
along with other observations, including TESS transits, 
to detect transit timing variations of the hot Jupiter WASP-4b.
The goal of the GMOS survey is to robustly estimate 
systematic effects and extract high-fidelity transit spectra for the target planets, 
ultimately constraining Rayleigh scattering and cloud deck prevalence in 
irradiated atmospheres. 
In this study, we focus on GMOS time-series transit spectroscopy 
observations of a well-studied hot Jupiter. 

HAT-P-1b is one of the first hot Jupiters discovered with the transit technique
\citep[][Table~\ref{tab:prop}]{bak07}.
The host star, HAT-P-1, is reasonably bright,
old \citep[3.6\,Gyr,][]{bak07}, and inactive \citep[e.g.,][]{knu10,nik14}. 
The equilibrium temperature of the planet's day side is approximately 1300\,K 
\citep[assuming full irradiation redistribution to the night side;][]{nik14}. 
The Rossiter-McLaughlin effect for the system was observed by \citet{joh08},
who showed that the planet's orbital axis is well aligned with the stellar
spin axis, indicating a post-formation orbital evolution that preserves alignment. 
Another aspect of the system is the presence of an equal-mass
stellar companion at an angular distance 
of 11.26\arcsec (V=9.75, F8V). 

\begin{NoHyper}
\begin{table*}[h]
\caption{Parameters for the HAT-P-1 system}
\begin{center}
\begin{threeparttable}
\begin{tabular}{lll}
    \toprule
M$_\star$ (M$_{\odot}$) & $1.15\pm0.05$ & \citet{nik14} \\
$\rm V_\star$ (mag)   &  $9.87\pm0.01$ & SIMBAD; \citet{zac13} \\
T$_{eff,\star}$ (K) &    $5980\pm50$ & \citet{nik14} \\
SpT$_\star$ & G0V & \citet{bak07} \\
L$_\star$ (L$_{\odot}$) & $1.59\pm0.10$ & \citet{nik14} \\
$\rm \log(g_\star$) & $4.36\pm0.01$ & \citet{nik14} \\\relax
[Fe/H]$_\star$ & $0.13\pm0.01$ & \citet{nik14} \\
Distance (pc) & $160\pm1$ & {\em GAIA} DR2 \\

\midrule
P (day) &  $4.46529976\pm(55)$ & \citet{nik14} \\
T$\rm _C$ (BJD$\rm _{TDB}$) & $2453979.93202\pm(24)$ & \citet{nik14} \\
a (au)  &  $0.0556\pm0.0008$ & \citet{nik14} \\
M$_p$ (M$\rm _J$) &   $0.525\pm0.02$ & \citet{nik14} \\
R$_p$ (R$\rm _J$)  &  $1.319\pm0.02$ & \citet{nik14} \\
T$_{eq,p}$ (K)  &   $1320\pm20$ & \citet{nik14} \\
inclination ($^\circ$) &  $85.63\pm0.06$ & \citet{nik14}\\

\bottomrule

\end{tabular}
\begin{tablenotes}
\end{tablenotes}
\end{threeparttable}
\end{center}
\label{tab:prop}
\end{table*}
\end{NoHyper}

The atmosphere of HAT-P-1b has been studied extensively in the past. \citet{tod10} observed
the planet's secondary eclipses using broadband photometry at 3.6, 4.5, 5.8, and 8.0\,$\mu$m
with the IRAC instrument on {\em Spitzer} and concluded that
the mid-infrared spectral energy distribution of the
planet's day side is consistent with a black body. The secondary eclipse of the planet was also detected in the K$\rm _s$ band using
the Long-slit Intermediate Resolution Infrared Spectrograph (LIRIS) on the William Herschel Telescope (WHT) \citep{moo11}.
In the following years, investigators focused on transit spectroscopy studies
using the {\em Hubble Space Telescope} (HST) with the 
Wide Field Camera 3 (WFC3) and the 
Space Telescope Imaging Spectrograph (STIS) instruments. 
\citet{wak13} used the G141 grism on the WFC3 to create a low-resolution
($\rm R \sim 70$)
transit spectrum in the wavelength range between 1.09 and 1.68\,$\mu$m.
These authors detected water absorption from the 1.4\,$\mu$m at more
than 5$\sigma$.
\citet{nik14} used HST/STIS to extract a low-resolution
transit spectrum between 0.29 and 1.03\,$\mu$m. These observations
yielded a detection of sodium at 0.589\,$\mu$m at 3.3$\sigma$.
\citet{sin16} summarized the HST observations and placed them in the context of
Spitzer transit photometry observations at 3.6 and 4.5\,$\mu$m,
and of other HST transit spectroscopy studies of hot Jupiters.
\citet{mon15} presented a very low-resolution transit spectrum in 
the visible range obtained with the Device Optimized for the Low Resolution (DOLORES) on the 3.6\,m
ground-based Telescopio Nazionale Galileo (TNG). Potassium has 
been detected in the atmosphere of HAT-P-1b in transit
using the Optical System for Imaging and low Resolution Integrated Spectroscopy (OSIRIS) on the Gran Telescopio Canarias (GTC) in tunable filter imaging mode
near 766.5\,nm \citep{wil15}.

In this work, we measure the optical transit spectra
we have obtained between 0.3 and 0.9\,$\mu$m,
using the GMOS-N instrument on the Gemini North telescope
and discuss them in the context of previous HST/STIS and 
TNG/DOLORES results in the visible range. 
In Section~\ref{sec:obs} we present the our observational strategy
and discuss the specifics of our data. Section~\ref{sec:red} details
our data reduction approach. Section~\ref{sec:fitting} focuses on 
our treatment of the time series and the transit spectrum extraction.
Section~\ref{sec:mod} discusses our data
modeling and the implications of our results.

%__________________________________________________________________
%__________________________________________________________________
%__________________________________________________________________
%__________________________________________________________________
\section{Observations}
\label{sec:obs}
\subsection{GMOS transit spectroscopy}
We observed two primary transits of HAT-P-1b using the Gemini Multi-Object Spectrograph
(GMOS-N) instrument on the Gemini North telescope (Maunakea, Hawaii) on 
2012 November 12 and on 2015 November 02 (Table~\ref{tab:data}).
Our observing strategy is similar to that of \citet{hui17} 
and to those of other earlier studies
using this instrument \citep[e.g.,][]{gib13a, gib13b, ste14, ves17}.

\begin{table*}
\caption{Observations}
\begin{center}
\begin{threeparttable}
\begin{tabular}{lllllll}
\toprule
Observation ID & Date & Band & Grating & No. of    & Duration & Airmass \\
               &      & (nm) &         & Exposures &          & range   \\
\hline\\[-1.5ex]
GN-2012B-Q-67 & 2012-Nov-12 & $520-950$ & R150 + G5306 &1018& 6h 13m & $1.06-1.98$ \\
GN-2015B-LP-3 & 2015-Nov-02 & $320-800$ & B600 + G5307 &378 & 5h 44m & $1.06-2.74$ \\
\bottomrule
\end{tabular}
\begin{tablenotes}
\end{tablenotes}
\end{threeparttable}
\end{center}
\label{tab:data}
\end{table*}

The goal of the observations was to perform accurate spectrophotometry of
the target star (HAT-P-1, or BD+37 4734B). Therefore, we configured GMOS
in its multi-object spectrograph mode (MOS) to observe the target and a suitable
reference star (BD+37 4734A, the binary companion to the target, 11.26$\arcsec$
separation). The target and reference stars have comparable spectral types and magnitudes
\citep[G0V with $\rm V=9.87$, and F8V with $\rm V=9.75$, respectively;][]{hog00, zac13}.

Both stars were observed through the same 30$\arcsec$ long slit because they 
are close enough that a single slit provides sufficient
sky coverage to assess the background. To mitigate any time-variable slit losses
that could complicate our time-series analysis,
we chose to use a wide slit (10\arcsec).
We ensured that the wavelength coverage was consistent between stars 
by selecting the position angle 
of the MOS slit mask to be equal to the PA
between the two stars ($74.3^{\circ}$ E of N).

The transit observation in 2012 was performed with the R150 + G5306 grating
combination (from now on, ``the R150 transit''). 
The OG515\_G0306 filter was used to block light at wavelengths shorter
than $\sim$520\,nm, as well as to block out contamination from other orders.
This yielded a wavelength coverage of 520 to 950\,nm.
The 2015 transit observation used the B600 + G5307 grating (the ``B600 transit''), with
a wavelength coverage of 320 -- 800\,nm. No blocking filter was used.

The ideal resolving power of the R150 observation is $\rm R=631$
at the blaze wavelength (717\,nm), and for the B600 observation, it is
$\rm R=1688$ at a blaze wavelength of 461\,nm. These values assume a
0.5$\arcsec$ wide slit \citep[GMOS Online Manual]{hoo04}, 
but our observations
use a 10$\arcsec$ wide slit and are seeing limited. 
The wavelength resolution we obtain is $2-6\times$
lower than the ideal values.

Our data were observed while GMOS was still operating with its
e2v DD detectors. These are decommissioned in GMOS-N since February 2017.
We reduced the read-out noise by windowing a region of interest
(ROI) on the detector in covering both spectral traces, rather
than reading the whole detector. A single ROI was used to cover both
stellar spectra. In addition, we binned the detector output ($1\times2$)
in the cross-dispersion direction. The detector was read out with six amplifiers and
with gains of approximately 2 e$^{-}$/ADU in each amplifier (although
the R150 spectral traces only cover four amplifiers).
Exposure times were chosen to keep count levels between $\sim$10,000 and
40,000 peak ADU and well within the linear regime of
the CCDs ($<50\%$ of the full well of the detector). 
The total duration of the transit observations was 6\,h 13\,m (R150)
and 5\,h 44\,m (B600). The transit duration of HAT-P-1b is 2\,h 40\,m, 
allowing sufficient out-of-transit sampling of the light curves.

\subsection{Long-term photometric monitoring of HAT-P-1}
\label{sec:lcodata}
We obtained photometric observations in the 
Johnson-Cousins/Bessell B-band of HAT-P-1b using the 
Las Cumbres Observatory (LCO) global network of robotic
telescopes \citep{bro13}. The network consists of 42 telescopes 
with mirrors with diameters of 40\,cm, 1\,m, and 2\,m. The telescopes are
spread across Earth in latitude and longitude, providing full sky coverage.
Our 674 photometric observations cover eight months between 2016 April 25 
and 2016 December 12 (one measurement every 8 hours on average, 
weather permitting, with flexible scheduling). We used 
only the 40\,cm and 1\,m telescopes with the SBIG 4k$\times$4k imagers.

\section{Data reduction}
\label{sec:red}
We based our data reduction on the custom GMOS transit
spectroscopy pipeline discussed in detail in \citet{hui17}. We used this code to correct
the raw images for gain and bias levels, to perform cosmic ray identification, to remove
bad columns, to flat-field the data, and to correct for slit tilt (as needed),
and finally, to extract the 1D spectra. We then applied corrections for
additional time- and wavelength-dependent 
dispersion shifts between target and
reference stars on the detector due to weather and airmass, for example. 
In this section we outline the main points of the pipeline 
and the additional corrections required by our data.

\subsection{Extracting the time series}
Cosmic rays were detected with 5$\sigma$ clipping using a moving boxcar
of 20 frames in time,
where the value of a given pixel was compared to the values of the same
pixel in the images obtained immediately before and after. The
value of the flagged pixel was replaced by the median value of that
pixel within the boxcar. The cosmic-ray removal flagged
several percent of pixels per image frame.

The GMOS-N e2v detectors suffered from bad
pixel columns, which we identified and excluded from our analysis. The pipeline
flagged 1.9-2.6\% of columns as bad by comparing the out-of-slit counts
within a given column with those of neighboring columns.
The majority ($\sim90\%$) of flagged columns are consistent between transits
and typically occur in the transition regions between detector amplifiers.
We saw no columns of shifted charge like those in the GMOS detector on the 
Gemini South telescope prior to July 2014, as reported by \citet{hui17}.

Flat-fielding should  ideally not be necessary because our measurement
is relative and we followed the counts in the same set of pixels as a function of time.
However, as the pointing changed throughout the night, the gravity vector on the instrument evolved, 
causing flexing. As a result, the spectral trace drifted over
different detector pixels during the observation. Therefore, we tested our extraction with and without
flat-fielding. We find that flat-fielding does not significantly affect the
scatter of the resulting white-light curves, increasing it (B600) or
decreasing it (R150) by several percent. For consistency, we chose to omit
flat-fielding in both observations. 

The HAT-P-1 transit spectrum observations do not suffer from spectral tilt. The sky
lines on the image are parallel to the pixel columns. We therefore did not correct for it.

We applied the optimal extraction algorithm \citep{hor86} to obtain the 1D spectra.
We tested a range of aperture sizes between 10 and 60 pixels in the cross-dispersion
direction. We find that the lowest scatter in the out-of-transit light curves
is produced by an aperture of 20 pixels for both observations.

\subsubsection{Slit losses}
Slit losses are caused by a combination of differential atmospheric dispersion and differential refraction
\citep[e.g.,][]{san14}. For the airmass ranges in our study (Table~\ref{tab:data}), a separation of 11.26",
and using Equation 3 in \citet{san14},
we estimate that the apparent shift in position due to differential refraction 
is between 0.002$\arcsec$ and 0.009$\arcsec$ (R150), and 0.002$\arcsec$ and 0.016$\arcsec$ (B600). This is much smaller than our 10" slit, and so it is unlikely that differential refraction plays a large role in our
observations. Because the reference and target stars form an almost equal-mass binary, they have very similar spectral
types, colors, and apparent brightness. Thus, differential
atmospheric dispersion is also unlikely to play a large role for our observations. 
Any variations are likely to be gradual with time and would be accounted 
for by our instrumental effects corrections (e.g., Section~\ref{sec:lcsys}, Equation~\ref{eqn:ramp}).

Another aspect is that any time jitter in the position angle (PA) of the slit could introduce differential slit losses.
No such variability is recorded in the FITS file headers, but low-level pointing changes cannot be completely
ruled out. To estimate the magnitude of this effect, we measured the exposure-to-exposure pointing jitter in the cross-dispersion direction and found that it is typically less than 0.04$\arcsec$ in both transit observations, 
compared to the 10\arcsec slit. The change in PA of the slit is likely of similar or smaller magnitude. 
Any slit losses that do occur would contribute to the red noise we observe in our light curves, but 
should at least in part be accounted for by our common mode correction (Section~\ref{sec:spec}).

\subsubsection{Diluted light}
\begin{figure*}
  \centering
  \includegraphics[scale=0.5]{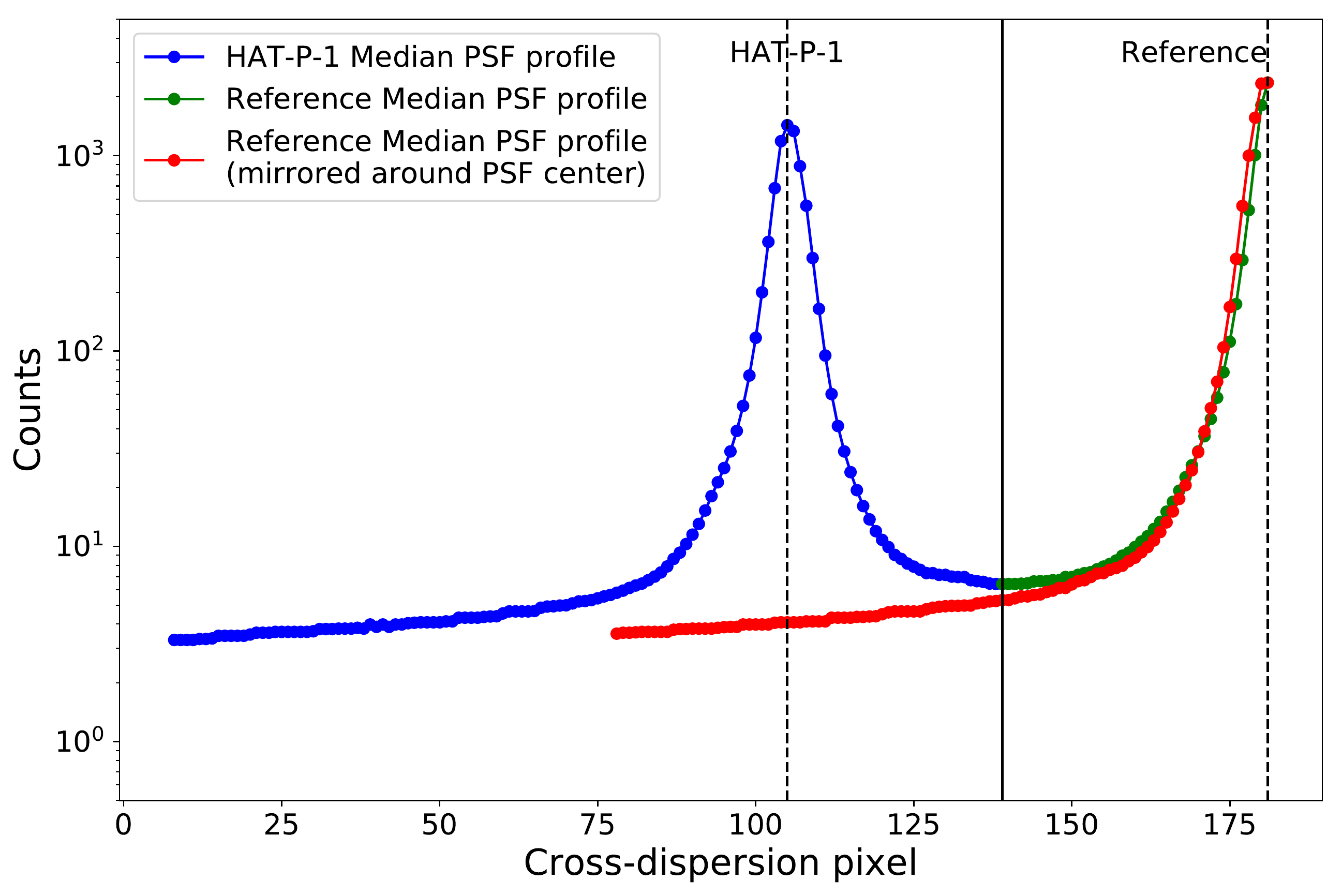}
  \caption{Median PSF profiles of the target and the reference star from the R150 data. The B600 observations look similar. We have mirrored the reference star
    PSF profile around its peak (green and red) in order to illustrate our method of correcting for
    diluted light. The black line indicates the point where we separate the two spectra. 
    Because we have ensured that the single-star PSFs observed with GMOS are symmetric, we can
    use the mirrored part of the PSF to estimate the amount of contamination on the target PSF. We
    also correct the reference star for diluted light from the target in a similar way. We extrapolate
    the red curve linearly for regions beyond the spatial range of the observation. 
    The boundary (black line) we select between the target and reference stars 
    was fixed as a function of time, although the spectral trace evolves with airmass and gravity vector
    on the instrument. However, this can be arbitrary 
    (as long as it is reasonable) because the exact location of the boundary does not affect 
    the amount of scattered light from the reference diluting the target flux extracted from 
    the core of the target PSF. This is measured only using the opposite wing of the reference PSF, which 
    is not affected by our choice of border. 
  }
  \label{fig:diluted_psf}
\end{figure*}

The reference star we used in our study is also the stellar companion of HAT-P-1 (11.2\arcsec)
that shares a similar magnitude and spectral type (planet host: G0, $\rm V=9.87$, companion: F8, V=9.75).
Because the two objects have a small separation on the sky, the reference star point spread function
(PSF) dilutes the transit light-curve measurement.
In order to estimate the magnitude of this effect, we measured the flux of the reference star at 11.2"
in the cross-dispersion direction opposite from the target. This flux is a combination of sky background
and scattered reference star light. We assumed that the sky background is spatially uniform within our
field of view, and we estimated its value by measuring the flux value farthest from the reference
star within the ROI in Figure~\ref{fig:diluted_psf}. The combined background and diluted-light fluxes
correspond to $\sim1\%$ of the target stellar flux.
To ensure that using the PSF shape on the opposite side of the target
is reasonable, we tested the symmetry of the GMOS spectral PSF
for a single star using the observations of WASP-4 \citep{hui17}. We found that at 11.2", the spectroscopic PSF is symmetric in the cross-dispersion 
direction within 0.2-0.4\% of the overall flux level at this location.

We caution that this needs to be accounted for when the robustness of the background measurement is tested.
One way to test the background estimation is to multiply the estimated background level in
a cross-dispersion slice by a constant factor, $f$, and test the effect of deliberately
over- or undersubtracting the background on the final results.
This test is not appropriate, however, if diluted light from a nearby companion is present 
because multiplying the background also increases the
slope of the diluting PSF that rises toward the reference star and does not take this into account in the optimal extraction stage. This increases
the wavelength-dependent dilution by
a factor $f$, but this increase is later not normalized by
dividing by a reference star flux that is also $f$ times higher. This can
cause spurious changes in the transit spectrum 
that would incorrectly suggest that the background subtraction is not reliable.

In our case, the reference and target
stars are clearly separated enough to ensure that the spatial slope of the diluting flux is 
approximately linear. We estimated the dilution strength by comparing 
the wavelength-dependent estimate of the dilution to the flux of the target star. 
The contamination level in the R150 data set of the target PSF caused by the reference PSF
is 0.1\% (near $\sim 500$\,nm) and increases to $\sim1.5$\% ($\sim 800$\,nm). 
In the B600 data, the dilution is $\sim1$\% and is stable in time and wavelength, except near the edges
of the observed wavelength range.
These dilution levels reduce the transit depth by $\sim10-200$\, 
parts per million (ppm), which corresponds to $\delta R_p/R_\star \lessapprox 0.015$, and this can be 
wavelength dependent. 

The dilution contamination increases to several percent near 
the edges of the observed wavelength ranges for both observations. This is particularly 
important in the bluest region of B600 ($\lessapprox 400$\,nm), but 
we discarded these data from our final analysis due to additional strong systematic
effects toward the end of the observation caused by high airmass.
The reddest part of the R150 observation ($\gtrapprox 850$\,nm) is also affected by strong dilution of the target star (3-4\% level). We do not use these data in Section~\ref{sec:mod} either. 

In order to test the effect of the dilution correction on the extracted planetary transit spectrum, 
 as an experiment, we performed our analysis with and without correcting the stellar spectra for dilution.
We find that in the B600 data, the effect of dilution on the 
measured transit depth is small, typically much smaller than 1$\sigma$ at a given wavelength. 
For R150, the uncorrected time-series spectra result in a slope in the
transit spectrum with shallower transits at longer wavelengths, 
where the dilution increases with wavelength, as a fraction of 
the total flux at R150, while at B600, the dilution ratio 
remains fairly constant with wavelength. In our final analysis, 
we therefore applied dilution correction to the R150 data but not to the B600 data. 

\subsubsection{Ghost spectra}
During both transit observations (B600 and R150), 
each spectral trace image was accompanied by a fainter trace-like signal.
The amplitude of the signal is small, $\sim50$\,ppm, compared to the
amplitude of the observed stellar spectrum. For both the reference
and the targets stars, it was confined to between 22 and 30 pixels from the
peak of the main trace, and followed the same spatial behavior as the
main trace in all wavelengths. Despite its low amplitude, this
spurious signal is clearly detected due to the extremely high
signal-to-noise ratio of the median-combined spectral image 
of the HAT-P-1 stellar system over the
entire night. However, attempting to
extract its spectrum is difficult because it is 
dominated by the wings of the PSF of the main spectral trace.
Thus, we cannot explore its spectral signature in detail, but
its persistent appearance in the same location in all images and
above both the target and reference stars suggests that it is
likely caused by internal reflections within the instrument.
Because it is highly localized, we masked the strip of pixels
that showed this spurious signal and did not use them in our analysis.

\subsection{Effect of terrestrial wind on the time series}
\label{sec:wind}
\begin{figure*}
  \centering
  \includegraphics[scale=0.5]{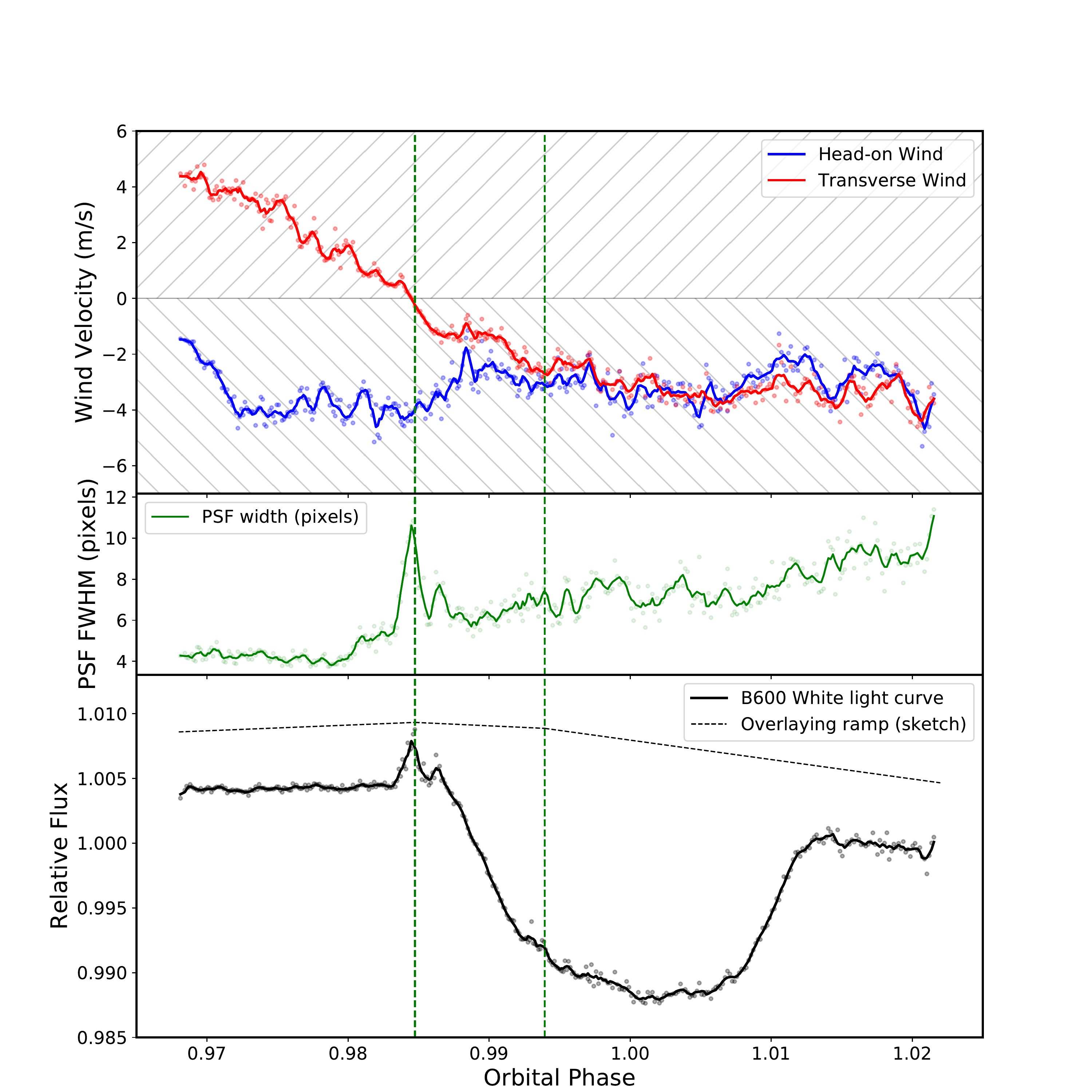}
  \caption{Top: Wind velocity and direction during 
  the B600 transit observation, with respect to the current pointing vector
  of the telescope. 
  The transverse wind component (red) comes from the right when positive 
  and from the left when negative in the observer reference frame because it follows the target star during the night. The head-on wind component (blue) comes 
  from the observer's back when negative and from the front when positive. 
  The transverse wind component coming from the right drops to zero near phase of 0.987 
  and then again increases in strength, but from the other direction.
  The head-on component of the wind
  remains fairly constant throughout the observation with respect to the
  dome opening, and always returns from the back of the observer.
  Middle: Median PSF FWHM for each exposure throughout
  the observation, showing a spike of a factor of $\sim$2 near phase 0.987, 
  which coincides with the change in direction in the transverse wind component. The overall 
  gradual increase in the FWHM beyond this point is likely caused by worsening seeing 
  caused by the increasing airmass of the observation. 
  Bottom: B600 white-light curve (black solid line) and a sketch of the ramp
  we use to correct the overall slopes.
  The green dotted vertical lines represent the two defects in the light curve 
  that appear to cause changes in the baseline slope. The defect near 0.987
  is likely induced by the change in the wind direction, while the cause of the
  other remains unclear. }
  \label{fig:winds}
\end{figure*}

The B600 white-light curve exhibits several unusual artifacts that could
not be corrected or correlated with any of the methods that are
frequently discussed in the literature. We explored
the headers of the raw fits file and discovered that one of these artifacts 
(near phase 0.987) coincides with a sudden change in the transverse wind
at the observatory site: the projected wind component, 
perpendicular to the line of sight of the telescope, 
switches direction from the right to the left
(Figure~\ref{fig:winds}). We suggest that the abrupt change in the 
apparent baseline slope of the B600 white-light curve near phase 
0.987 could be due to structural vibrations caused by the change in wind
direction with respect to the line of sight, which affect the spectral PSF width. 
The timing of this defect is unfortunate because it coincides with the ingress of the transit. 
We removed these data from our analysis. 

The leading one hour of the R150 white-light curve ends with a kink near phase 0.9805. 
At about this time, the direction of the head-on component of 
the wind from changes. It comes from the front instead of the back, as before. In this data set, 
the transverse component of the wind always comes from the left of the observer
and is approximately constant. This change in the wind does not produce a corresponding spike 
in the spectral PSF full width at half-maximum (FWHM), as in the B600 observation, and we cannot unambiguously 
attribute the kink to it. 

\subsection{Wavelength calibration}
After spectral extraction, we obtained the wavelength solution
using CuAr lamp spectra taken on the same
day as each science observation. To observe the CuAr
spectra at high resolution, we used a separate MOS slit mask
to that used for science, with the same slit position
and slit length as the science mask, but the slit width was
only 1\arcsec (compared to 10\arcsec).
The arc frames were obtained with
the same grating and filter setup as the corresponding science observation.

We used the {\it identify} task in the Gemini IRAF package to identify
the spectral features in the CuAr spectra.
A wavelength solution was then constructed by fitting
a straight line to the pixel locations of the spectral features
and their known wavelengths. This solution was then refined
by cross-correlating all spectra with the pixel 
locations of several known stellar and telluric features.

The final uncertainties in the wavelength solution are
approximately 1\,nm for all observations, which is 
$\sim3\%$ and $\sim6-7$\% of the bin widths 
used in the final transmission spectrum for the R150 and B600 transits, 
respectively.
This level of uncertainty in the wavelength solution has been shown
to be sufficient to avoid systematic effects caused by
wavelength-dependent differences in the limb-darkening models
we used to fit the transits in Sect.~\ref{sec:fitting} \citep{hui17}, 
and in addition, it is also smaller than our resolution element ($\sim4$\,nm in R150 
and $\sim$2\,nm in B600).

\subsection{Dispersion-direction shifts of the stellar spectra}
Because GMOS has no atmospheric dispersion compensator, we expect
the wavelength solution to shift with time.
\citet{hui17} reported that throughout the night, individual
spectra are shifted and stretched in the dispersion direction.
The shift and stretch are both time and wavelength dependent.
Left uncorrected for, this effect could introduce spurious signals
in the observed transit spectrum of the planet.
In order to avoid these systematic effects in the final light curves
related to shifting wavelength solution with time, we applied
a correction for this effect.

A priori differential refraction calculations 
alone do not explain the observed shifts for our
observations well. Therefore we applied a cross-correlation to four spectral segments
as a function of time to measure the spectral shift and stretch
empirically, similarly to \citet{hui17}.

Finally, we corrected for the differential offset
in wavelength between the reference and target star spectra
on the detector. This offset is caused by the fact that the PA of the
detector slit is slightly different from the PA of the reference star.
We used a cross-correlation to measure the offset ($1.57 - 1.63$\,
pixels for all observations). We interpolated the reference star’s
spectrum onto the target star’s wavelength solution.
Bad columns (which are the same columns on the detector but are at
different wavelengths for each star) were omitted.

We tested the effect of wavelength shifts with time by running our
analysis with and without the correction. We find that for the 
low-resolution broad bins that we adopted, the correction results 
in differences typically
smaller than 0.1\,$\sigma$ in both B600 and R150. However,
the shift correction has a small effect on the narrow bins
we used to investigate the Na absorption region. Therefore we
chose to keep this correction in our analysis. 

\section{Transit light-curve analysis}
\label{sec:fitting}

We describe our transit depth measurements for the 
B600 and R150 observations below. First, we describe 
the model we used to correct for the systematic effects in Section~\ref{sec:lcsys}. 
In Section~\ref{sec:wlc} we focus on 
the wavelength-integrated white-light curves for each of the observations. 
Section~\ref{sec:spec} focuses on the transit light curves in individual
wavelength bins, while Section~\ref{sec:res_sodium} discusses the time series
near the 589\,nm Na I resonant doublet expected in the planet's atmosphere.
Section~\ref{sec:starvar} addresses any potential variability in the host
star based on the LCO data. 

\subsection{Light-curve model of systematic effects}
\label{sec:lcsys}
To describe the systematic effects (telluric absorption and instrumental effects)
in our light curves, we adopted the parameterization of \citet{ste14}, 
who used the same instrument, GMOS, to study the transmission 
spectrum of the hot Jupiter WASP-12b in the red-optical. 
This model addresses effects caused by the Cassegrain Rotator  Position  Angle (CRPA), 
the airmass, and the time-varying PSF of the instrument. 
Our light-curve model also accounts for any approximately 
linear or quadratic trends as a function of time that are 
commonly seen in transit studies. We applied this model to both the R150 and B600 data 
and to both the white-light curves and the wavelength-dependent light curves. 

We parameterized the systematic effect due to the CRPA
by approximating the photometric dependence of the
flux on the CRPA as a cosine function:
\begin{equation}
  S(A_{CRPA}, \theta_{CRPA}) = 1 + A_{CRPA}\cos(\theta(t) + \theta_{CRPA}).
\end{equation}
Here, $A_{CRPA}$ and $\theta_{CRPA}$ are the free parameters, while $\rm \theta$(t) is
the known CRPA as a function of time. Like \citet{ste14}, we considered a
systematic trend, or a "ramp", as a function of time in the light curve in the form, 
\begin{equation}
  R(t) = c_{ramp} + b_{ramp}(t-t_0) + a_{ramp}(t-t_0)^2,
  \label{eqn:ramp}
\end{equation}
where $a$ and $b$ are free parameters, while $t$ is
time and $t_0$ is the time of observation of the first frame in the time series.
However, we find that with our data, $a_{ramp}$ is degenerate with the 
transit depth, and based on the resulting Bayesian information criterion (BIC)
values, the data are better described by a pure linear ramp. We included an airmass correction: 
\begin{equation}
  A(\alpha) = C_{\alpha}\alpha(t),
\end{equation}
where $\alpha(t)$ is the known run of airmasses at the time of observation, and 
$C_{\alpha}$ is a free parameter. 

These parameters account well for the long-timescale (hours) variations of the light curves, 
but do not correct short-timescale ($\lesssim10$\,min) effects. To address these, 
we experimented with several time-varying  
quantities: the ratio of the widths of the target and reference PSFs, the cross-dispersion
position of the target PSF on the detector, and the dispersion direction shifts of the target spectrum.
We find that most of these are correlated with the light curves at some level, but
the data are best described by the width of the target PSF as a function of time: 
\begin{equation}
  P(W) = C_{PSF\,width}W(t),
\end{equation}
where $C_{PSF\,width}$ is a free parameter, and W(t) is the run of the median PSF widths throughout
the observations. This correction term is likely related to the state of the atmosphere above the telescope 
during the observations.

Thus, the final model to fit to the light curve (normalized to 
the out-of-transit flux) can be represented as
\begin{equation}
  F(t) = T(t) S(A_{CRPA}, \theta_{CRPA}) R(t) A(\alpha) P(W).
\end{equation}
$T(t)$ is the astrophysical transit light-curve model,
normalized to 1 out-of-transit.

We used several open-source packages to model and fit
the transit light curves. $T(t)$ was calculated with
the transit code {\emph{batman}} \citep{kre15}, and for the fit we used
{\emph{emcee}}, a pure-Python implementation of the
affine-invariant Markov chain Monte Carlo (MCMC) ensemble sampler
\citep{goo10, for13}. The stellar limb darkening was pre-computed
for each wavelength bin using PyLDTk \citep{par15}, which uses the spectral library in \citet{hus13}, based on the
atmospheric code PHOENIX. We used the nonlinear parameterization 
for limb-darkening coefficients \citep{cla00,sin10}. 

\subsection{White-light curve}
\label{sec:wlc}
We constructed transit white-light curves for each transit (B600 and R150)
by integrating each observed target and reference spectrum over wavelength 
and dividing one by the other.
This approach eliminates many of the light-curve artifacts
caused by variations in Earth's atmosphere. 
Because some transit parameters are not wavelength dependent
(e.g., the ratio of semimajor axis
to stellar radius, $a/R_\star$, the central time of transit, $T_0$, and
the orbital period, $P$), we can use the precise
white-light curves, which have a high signal-to-noise ratio, 
to measure them, rather than the noisier light curves at specific wavelengths. 

\begin{figure*}[t]
  \centering
  \makebox[\textwidth][c]{\includegraphics[width=1.1\textwidth]{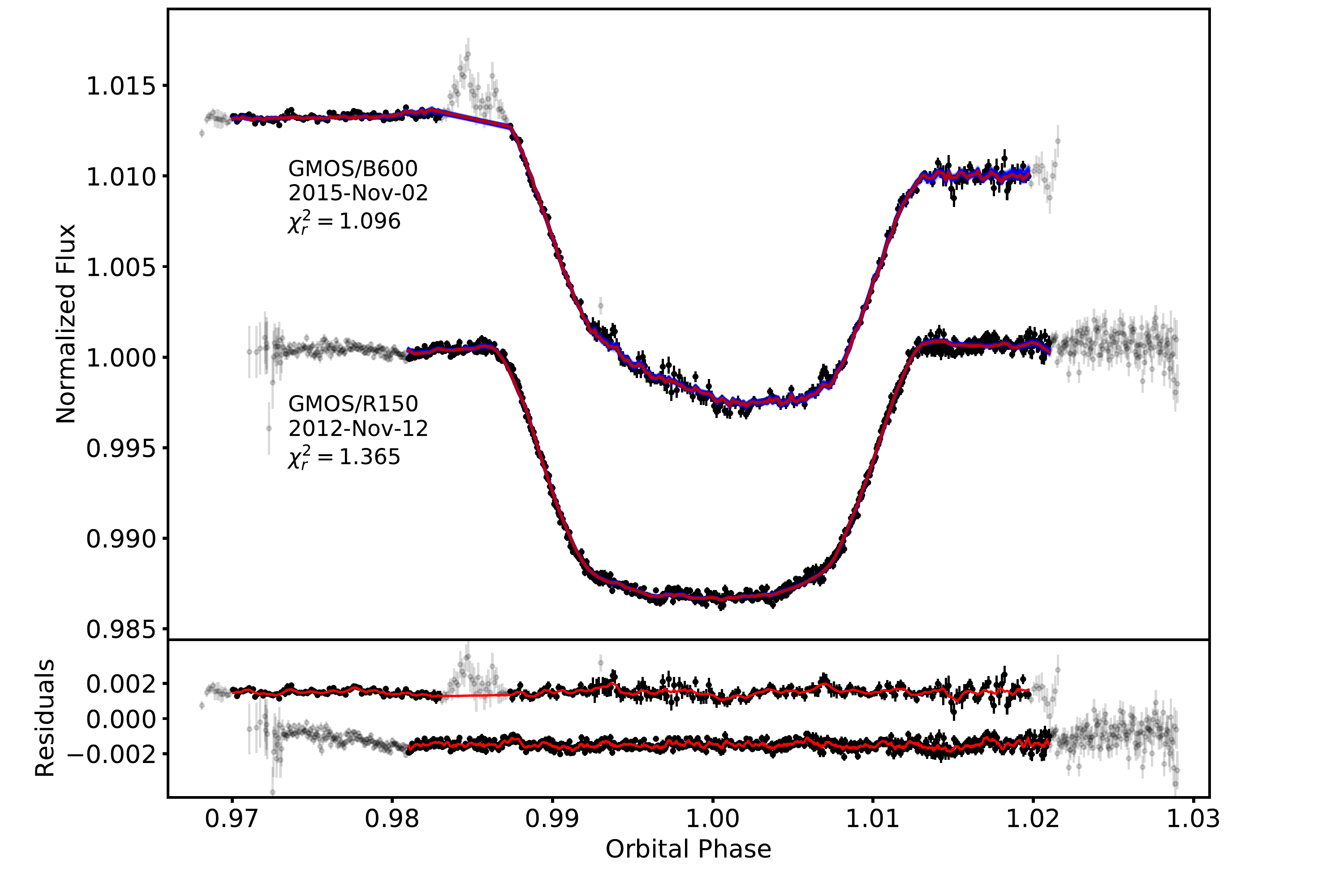}}
  %\vspace*{-10mm}
  \caption{White-light curves of the transits observed with the B600 and R150 
    grating spectral time series integrated over wavelength between 330 and 770\,nm
    and 550 and 950\,nm, respectively. Gray points indicate poor-quality data 
    excluded from the analysis. The anomalous peak near phase 0.987 in the B600 
    light curve has an amplitude of about 0.5\% and represents a
    defect in the light curve, likely caused by rapid change in the wind 
    at the observatory site, correlated with a spike in the PSF FWHM (Figure~\ref{fig:winds}). 
    Near phase 0.987, we exclude the part of the spectral time series with anomalously broad PSFs. 
    The R150 light curve appears to be better behaved 
    and displays no large-scale irregularities, but the initial hour of the light curve 
    is difficult to account for with our model of systematic effects. We drop it from the analysis, along with the final $\sim45$\,min, which were
    obtained at high airmass and have a large scatter. 
    The red lines represent the best-fit model
    of the transit and systematic effects for the white-light curve (top) and
    the smoothed residuals  to highlight any residual red noise (bottom).
    The faint blue lines in the top panel represent sample model instances
    from the converged MCMC chains. }
  \label{fig:wlc_fit_both}
\end{figure*}

The data cover wide wavelength ranges from the blue optical to the
near-infrared. 
Our B600 light curve includes intensity measurements between 330 to 770\,nm. 
Similarly, we integrated the R150 spectral time series between 550 and 950\,nm
(Figure~\ref{fig:wlc_fit_both}). We estimated the uncertainties of each photometric 
point in two ways: from photon noise (assuming the only 
source of uncertainty of a given point is due to Poisson noise of the incoming photons), 
and based on the local scatter of the points, 
after subtracting a white-light curve smoothed by convolving it with a Gaussian. The latter
approach is an empirical estimate of the uncertainty of the points, and 
we used it to fit the white-light curve. 

As we discuss in Section~\ref{sec:wind}, there is a sharp change in the slope of the
ramp in the R150 data (Figure~\ref{fig:wlc_fit_both}), possibly related to a change in 
the head-on component of the wind. However, phase 0.9805 is also 
close to the phase where 
the target crossed the meridian (airmass of 1.057), and telescope field rotation may also 
have contributed. Regardless, the residuals of the first hour of observations 
(Figure~\ref{fig:wlc_fit_both}, bottom panel) show a different slope than the 
rest of the observation. Because this happens well before egress, we 
simplified our analysis and dropped these data. The final 45\,min of R150 data were 
also excluded because they were taken at high airmass and their residuals are higher 
than those of the preceding white-light curve points. 

We also removed the data near phase 0.987 in the B600 data set, where the light curve is strongly 
distorted. We removed all phases covered by the unusual spike in the PSF FWHM, 
which coincides with a change in the direction from which the transverse wind component impacts the dome (Figure~\ref{fig:winds}). We also dropped the last several minutes of this light curve, taken at high airmass
where the white-light curve appears to become unstable (Figure~\ref{fig:wlc_fit_both}).

\begin{figure*}

  \makebox[\textwidth][c]{\includegraphics[width=1.1\textwidth]
                                                {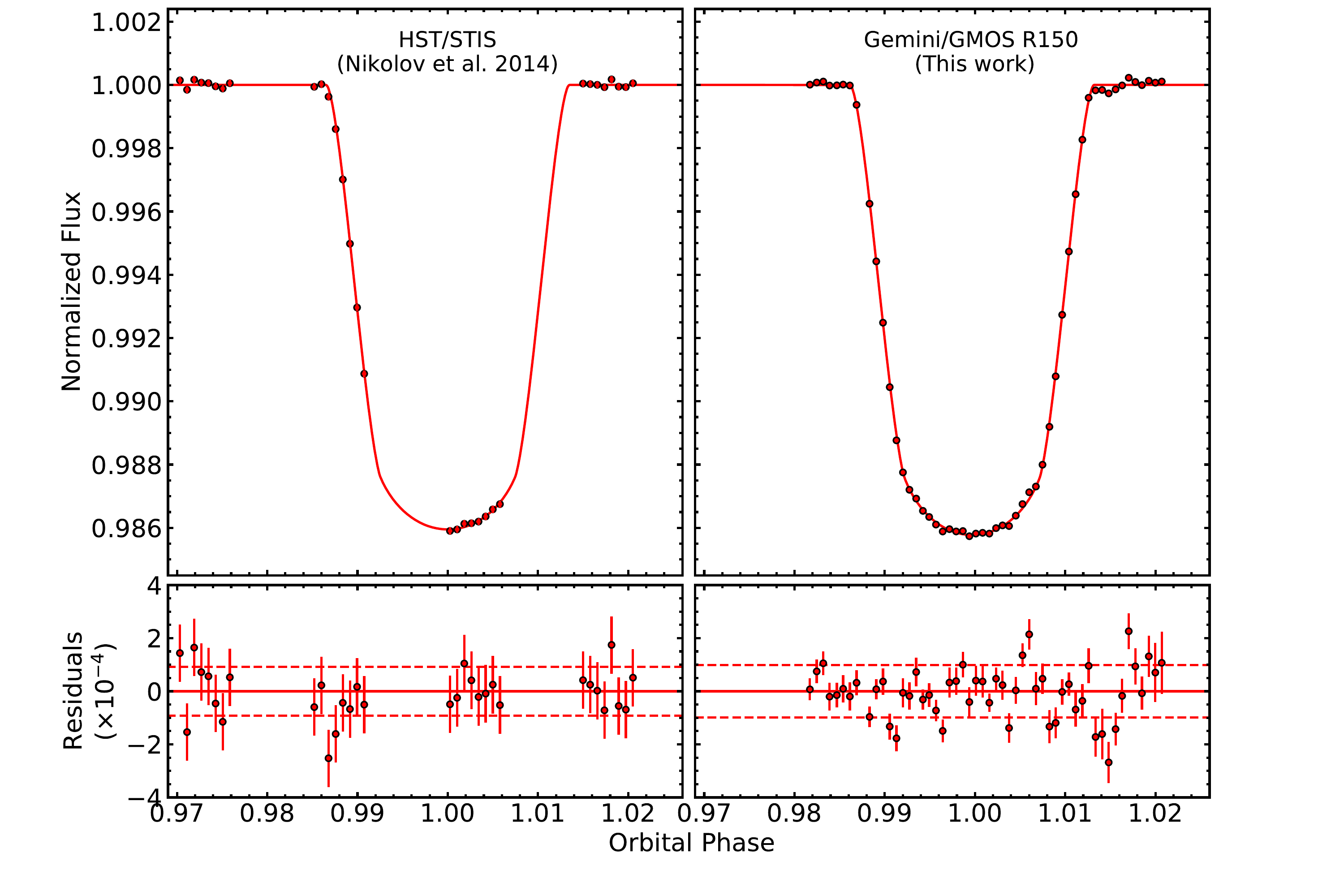}}
  %\vspace*{-10mm}
  \caption{
  We compare the GMOS R150 transit white-light curve (right panels, 4$\sigma$ outliers removed)
  after correction for systematic effects to a corrected HST/STIS optical white-light curve 
  \citep[left panels,][]{nik14}. 
  The HST transit was observed on 2012 May 30 (Visit 20) with the G750L grating (524 - 1027\,nm). 
  The GMOS R150 white-light curve is binned to match the HST integration time (284\,s). In the white-light curves, 
  the scatter of the residuals is comparable between the two instruments. 
  The rms of the residuals (dashed lines) is 90\,ppm (HST/STIS) and 100\,ppm (Gemini/GMOS), 
  but GMOS can observe the transit continuously, while HST observations 
  have gaps because the spacecraft orbits Earth. While the R150 observation is
  comparable in precision to that with HST/STIS, the B600 observation (not shown) is subject to more
  complex systematic effects and the corresponding rms of the residuals and the uncertainties
  in the transit parameters is higher than for space-based observations at similar wavelengths.
  }
  \label{fig:wlc_hst}
\end{figure*}

We fit the white-light curve with the model described 
in Section~\ref{sec:lcsys}. In the statistical fits, 
we allowed the semimajor axis ($a/R_{\star}$), 
the transit depth ($R_p/R_{\star}$), and the central time of transit 
($T_0$), which are a part of the $T(t)$ calculation, to vary freely. 
A large uncertainty in the orbital inclination could result in a degeneracy 
with the slope of the transit spectrum \citep{ale18}. 
However, the orbital parameters of 
HAT-P-1b are well known \citep[e.g.,][]{nik14}, therefore we 
fixed the limb-darkening coefficients and inclination of the planetary orbit 
in our analysis (Table~\ref{tab:fitting}). 

Unlike the R150 white-light curve, the B600 white-light curve is poorly described by a single linear ramp.
The white-light curve appears to increase slightly 
until the defect near orbital phase 0.987, after which 
the ramp slope appears to decrease with a shallow slope 
during ingress and then sharply drop off near
phase 0.994 (Figure~\ref{fig:winds}). 
We experimented with a segmented version of the linear ramp and found that 
the fit quality and the BIC are optimal for a ramp with three segments (before phase 0.987, between 0.987 and 0.994, and after 0.994), where each segment is linear with
a free slope. 
We explored a number of different configurations for the break 
points and correction parameters for the white-light curve. 
We tried to fit the B600 white-light curve without any break points, 
and using the airmass function to correct for the changing slope. 
This led to a reduced $\chi^2$ of the fit of 1.9 and $BIC=650$. 
Relying on a single break point improved these values, but the best fits were achieved 
with two break points: reduced $\chi^2 = 0.94$, and $BIC = 360$, showing a 
clear preference for the models with two break points. 
The location of the break near 0.994 is clearly visible by eye, 
while for the break near 0.987, we ran the white-light curve fit on a grid of locations 
and selected the location that yielded the best BIC. We experimented 
by letting the break-point locations vary as a free parameter, but the MCMC fits 
did not converge, even with strict priors. We therefore fixed the
break-point locations. At these points the adjacent segments of the ramp 
were required to have the same value to avoid discrete jumps in the ramp. 

Additional detrending options can be explored \citep[e.g.,][]{ste14}, 
but regardless of that choice, the ramp slope of this observation is variable 
and changes sharply during ingress and the transit itself. 
Because of this, any decorrelation function that could be applied to the B600 data 
would be degenerate with the transit depth, leading to larger
transit spectrum uncertainties (Section~\ref{sec:spec}).

We present the R150 white-light curve corrected for systematic effects in 
Figure~\ref{fig:wlc_hst} and compare it to an HST/STIS white-light curves in the optical \citep{nik14}. 
In our fit, we kept the inclination of the system fixed to the value found by \citet{nik14}, based
on multiple HST transits with the STIS and WFC3 instruments. 
The $a/R_{\star}$ parameter we find for the R150 light curve is well within 1\,$\sigma$
of the HST value.

\begin{table*}
\caption{Best-fit parameters for the white-light curve MCMC}
\begin{center}
\begin{threeparttable}
\begin{tabular}{lll}
\toprule
Parameter & B600 white-light curve & R150 white-light curve \\
\hline\\[-1.5ex]
Orbital parameters \\
\hline\\[-1.5ex]
T$_0$ ($\rm BJD_{TDB}$) & $2457328.907659 \pm (34)$ & $2456243.839084 \pm (76)$ \\
$R_p/R_{\star}$ & $0.1198 \pm 0.0010$ &  $0.11899 \pm 0.00035$  \\
$a/R_{\star}$ & $9.8775 \pm 0.0147$ & $9.8387 \pm 0.0056$\\
$i$ (degrees) & 85.634$^{\circ}$ (fixed) & 85.634$^{\circ}$ (fixed) \\
$e$ (eccentricity) & 0 (fixed) & 0 (fixed) \\

\bottomrule
\end{tabular}
\begin{tablenotes}
\end{tablenotes}
\end{threeparttable}
\end{center}
\label{tab:fitting}
\end{table*}

\subsection{Transit spectrum}
\label{sec:spec}
We divided the time-series spectra into wavelength bins. For consistency, 
we used 200\,pixel bins in both R150 and B600, which corresponds to bins
of 36\,nm and 16\,nm, respectively. To measure the transit spectrum of
HAT-P-1b, we applied the model discussed in the previous
section to the transit light curves corresponding to
individual wavelength bins ($\rm \lambda$LC) with several changes. 
To correct for wavelength-independent systematic effects
that were not accounted for by our white-light curve model, we divided
each $\lambda$LC by the white-light curve residuals (data minus best-fit model). 
We smoothed the white-light curve residuals by convolving them with a
Gaussian to reduce the effect of high-frequency noise 
when the light curve was divided. When we skipped this step, the point-to-point scatter 
in the R150 and B600 $\rm \lambda$LCs typically increased by $\sim20$\% and $\lesssim 5$\%, respectively. 
This common-mode correction resulted in lower $\chi^2$ fits, 
but did not constitute an additional free parameter because we always
divided the light curves by the same white-light curve residuals. 

We fixed the ratio of the semimajor axis to stellar radius, $a/R_\star$,
the central time of transit, $T_0$ to their best-fit values from 
the white-light curve fits because they are wavelength independent. The transit depth,
$A_{CRPA}$, $\theta_{CRPA}$, $C_{\alpha}$, and $C_{PSF\,width}$ remained free, as were the linear ramp parameters. As with the white-light curve, we used a segmented ramp for the 
B600 $\lambda$LCs. 

\begin{figure*}[t]
  \centering
  \makebox[\textwidth][c]{\includegraphics[width=1.0\textwidth]{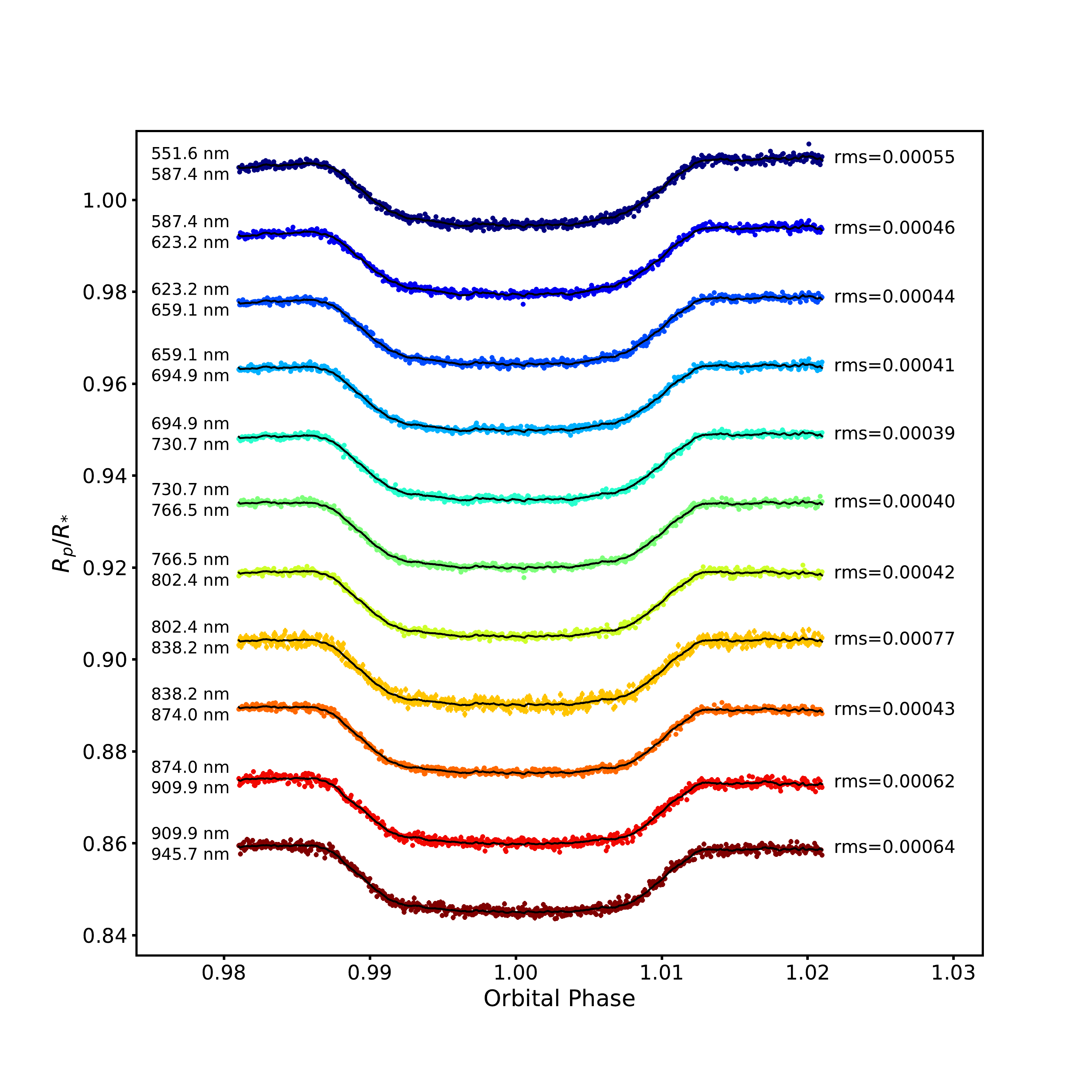}}
  %\vspace*{-15mm}
  \caption{R150 light curves in each wavelength bins (labeled) and 
    the best-fit transit models (black lines). We indicate the rms values
    of the residuals between observation and model 
    in order to quantify the scatter in each wavelength bin. 
  }
  \label{fig:r150_colors}
\end{figure*}

\begin{figure*}

  \makebox[\textwidth][c]{\includegraphics[width=1.0\textwidth]
                            {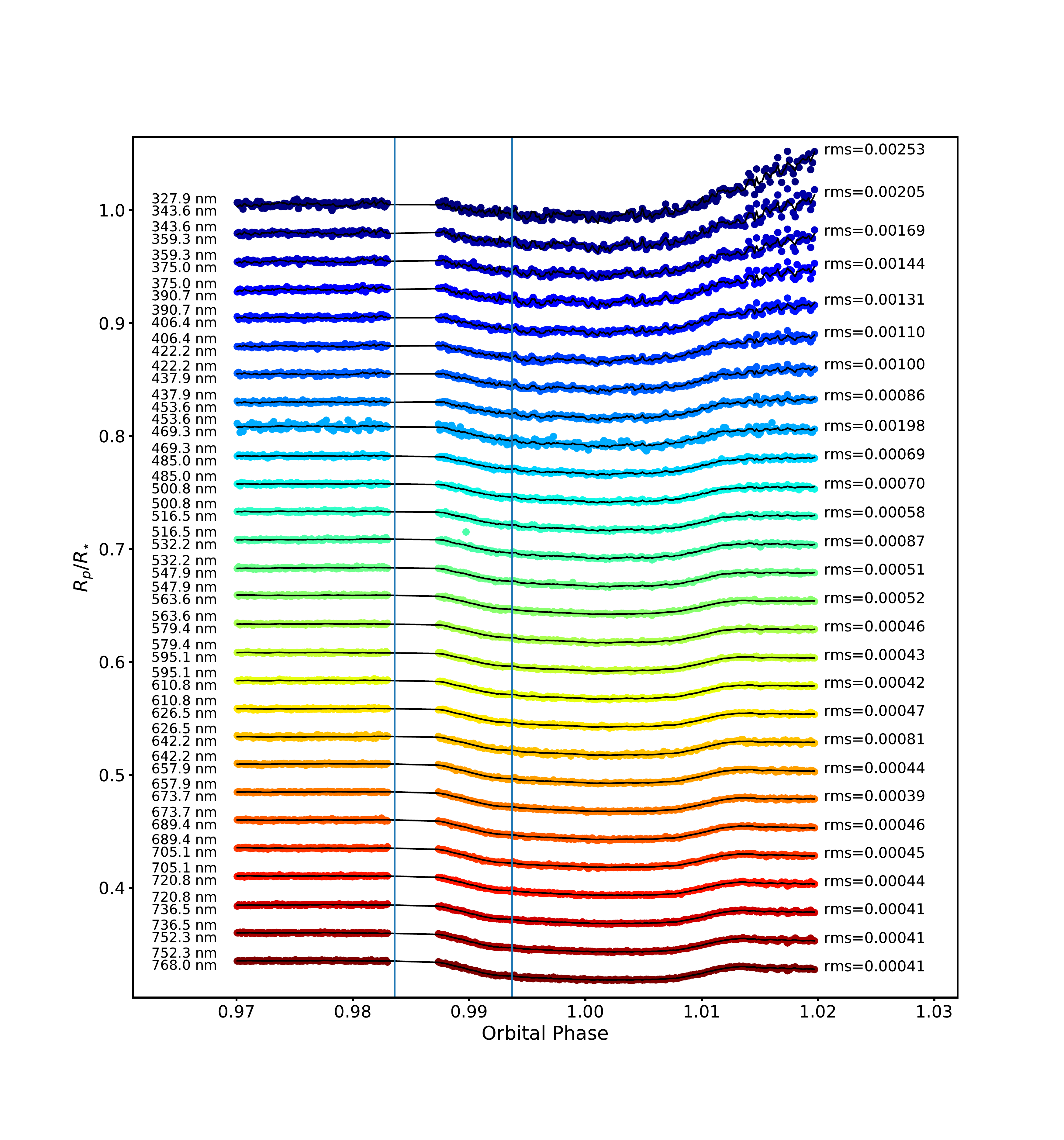}}%
  %\vspace*{-20mm}
  \caption{Similar to Figure~\ref{fig:r150_colors}, but for the B600 data. 
    We denote our transit model break points with vertical blue lines. 
  }
  \label{fig:b600_colors}
\end{figure*}

We fit each light curve in every bin again using the PyLDTk code 
to compute the limb-darkening coefficients and the \emph{emcee} 
package to retrieve the posterior distributions of the 
wavelength-dependent free parameters. We plot the transit fit results 
in Figures~\ref{fig:r150_colors} and \ref{fig:b600_colors}. 
While the typical R150 light curve appears to be well corrected 
for systematic effects, the fits near 750\,nm and $\gtrsim850$\,nm 
are potentially unreliable due additional
correlated noise effects of strong and variable telluric absorption
by O$_2$ and H$_2$O, respectively. 
The B600 data at wavelengths shorter than $\sim400$\,nm are heavily affected by 
high airmass toward the end of the observation. We tabulate 
both the R150 and B600 transit spectra in Table~\ref{tab:tr_spec}.

The uncertainties of photometric points in each light 
curve were estimated as described in Section~\ref{sec:wlc}. We estimated how well
our data corrections work by comparing uncertainties 
in $R_p/R_{\star}$ for the fits where we assumed pure photon noise for the 
points in the light curves to the fits using the 
empirical photometric scatter (Table~\ref{tab:tr_spec}). 

\begin{table}[h]
\caption{HAT-P-1b transit spectrum}
\begin{center}
\begin{threeparttable}
\begin{tabular}{cccc}
    \toprule

    Wavelength & R$_p$/R$\rm _\star$ & $\sigma/\sigma_{photon}$ $^1$ & Data \\
     (nm)      &                     &                               & quality$^2$ \\
    \midrule
    \underline{R150} & & \\
    551.6 - 587.4       & 0.11915 $\pm$ 0.00085 & 1.5 & 1 \\
    587.4 - 623.2       & 0.12088 $\pm$ 0.00064 & 1.4 & 1 \\
    623.2 - 659.1       & 0.12045 $\pm$ 0.00069 & 1.5 & 1 \\
    659.1 - 694.9       & 0.11919 $\pm$ 0.00056 & 1.5 & 1 \\
    694.9 - 730.7       & 0.11989 $\pm$ 0.00054 & 1.4 & 1 \\
    730.7 - 766.5       & 0.11876 $\pm$ 0.00059 & 1.5 & 0 \\
    766.5 - 802.4       & 0.12003 $\pm$ 0.00062 & 1.6 & 1 \\
    802.4 - 838.2       & 0.11865 $\pm$ 0.00110 & 2.7 & 1 \\
    838.2 - 874.0       & 0.11898 $\pm$ 0.00061 & 1.4 & 1 \\
    874.0 - 909.9       & 0.11793 $\pm$ 0.00087 & 1.7 & 0 \\
    909.9 - 945.7       & 0.11859 $\pm$ 0.00081 & 1.3 & 0 \\
    \midrule
    \underline{B600} & & \\
    327.9 - 343.6       &       0.09987 $\pm$   0.01503 & 4.1 & 0 \\
    343.6 - 359.3       &       0.10876 $\pm$   0.01066 & 2.4 & 0 \\
    359.3 - 375.0       &       0.10976 $\pm$   0.00900 & 4.3 & 0 \\
    375.0 - 390.7       &       0.11828 $\pm$   0.00707 & 3.6 & 0 \\
    390.7 - 406.4       &       0.11393 $\pm$   0.00517 & 2.3 & 1 \\
    406.4 - 422.2       &       0.11779 $\pm$   0.00450 & 2.7 & 1 \\
    422.2 - 437.9       &       0.11522 $\pm$   0.00374 & 2.2 & 1 \\
    437.9 - 453.6       &       0.11983 $\pm$   0.00331 & 1.2 & 1 \\
    453.6 - 469.3       &       0.12232 $\pm$   0.00577 & 4.0 & 1 \\
    469.3 - 485.0       &       0.11977 $\pm$   0.00231 & 1.7 & 1 \\
    485.0 - 500.8       &       0.11791 $\pm$   0.00223 & 1.7 & 1 \\
    500.8 - 516.4       &       0.12090 $\pm$   0.00174 & 1.6 & 1 \\
    516.4 - 532.2       &       0.12286 $\pm$   0.00276 & 2.7 & 1 \\
    532.2 - 547.9       &       0.12011 $\pm$   0.00177 & 1.0 & 1 \\
    547.9 - 563.6       &       0.11905 $\pm$   0.00177 & 1.7 & 1 \\
    563.6 - 579.4       &       0.12020 $\pm$   0.00160 & 1.7 & 1 \\
    579.4 - 595.1       &       0.12133 $\pm$   0.00152 & 1.9 & 1 \\ 
    595.1 - 610.8       &       0.12042 $\pm$   0.00169 & 1.8 & 1 \\ 
    610.8 - 626.5       &       0.12039 $\pm$   0.00163 & 2.0 & 1 \\
    626.5 - 642.2       &       0.12073 $\pm$   0.00277 & 3.3 & 1 \\
    642.2 - 657.9       &       0.12086 $\pm$   0.00163 & 2.0 & 1 \\
    657.9 - 673.7       &       0.12136 $\pm$   0.00157 & 1.6 & 1 \\
    673.7 - 689.4       &       0.12079 $\pm$   0.00162 & 1.5 & 1 \\
    689.4 - 705.1       &       0.12060 $\pm$   0.00177 & 2.2 & 1 \\
    705.1 - 720.8       &       0.12027 $\pm$   0.00178 & 2.4 & 1 \\
    720.8 - 736.6       &       0.11938 $\pm$   0.00145 & 1.8 & 1 \\
    736.6 - 752.2       &       0.11890 $\pm$   0.00179 & 2.4 & 1 \\
    752.2 - 768.0       &       0.12157 $\pm$   0.00162 & 2.3 & 1 \\
\bottomrule
\end{tabular}
\begin{tablenotes}
\item[1] The ratio between the uncertainty of $R_p/R_{\star}$
at a given wavelength, assuming a realistic estimate of the noise 
per spectrophotometric point, compared to an ideal photon noise case. 

\item[2] We present all transit depths we measured, but flag some of them 
with poor data quality (value of 0). These measurements were excluded 
from further analysis. Specifically, the bluest light curves in B600 are affected 
by high airmass at the end of the observation. The R150 data 
near 750\,nm and $>850$\,nm might be affected 
by O$_2$ and H$_2$O absorption bands, and we considered them unreliable.
Robust transit depth measurements are marked with a data quality value of 1.

\end{tablenotes}
\end{threeparttable}
\end{center}
\label{tab:tr_spec}
\end{table}

\subsection{Core of the Na I absorption line}
\label{sec:res_sodium}
Both the R150 and the B600 spectra cover the Na I doublet at 589\,nm. Sodium has
been detected in the atmosphere of HAT-P-1b by
%\LEt{to properly include these two references into the main text, please substitute "and" for the semicolon. This is a LaTeX command error that I cannot fix for you with the program I work with (citet and citep)} 
\citet{nik14} and \citet{sin16} in the narrow
core of the line (3nm wide bin centered on 589.3nm at $3.3\sigma$). However, these authors did not detect 
the pressure-broadened wings of the Na feature.

We explored this wavelength region by comparing the narrow-band transit depth in the core of the
sodium line in both the R150 and B600 observations
with the broadband transit depth in that region. We divided the region around the Na I 
line into narrow-band bins of 2\,nm each. The bins overlap each other by 1\,nm 
in order to ensure that any increase in measured $R_p/R_{\star}$ 
is not due to a single anomalous pixel or pixel column. 
The overlapping-bin light curves were fit as described in Section~\ref{sec:spec}.

If the feature is real, it should ideally result in a smooth rise and 
drop of transit depths as a function of wavelength, essentially tracing the dispersion 
function of the instrument. This approach does not rule out wavelength-dependent 
systematic effects, but prevents our results from being influenced by a spurious feature due to a pixel defect 
that might produce a sharp and discontinuous change in transit depth.
We tested this on wavelengths beyond the immediate vicinity of
the Na I feature where we expect the spectrum to be flat. Seeing features
in our spectrum in this region might indicate that narrow-band transit depth
variations are difficult to measure reliably with Gemini/GMOS. In 
addition, we experimented with the ``divide white'' approach described by \citet{ste14}, for example, where we compared the transit depth in the narrow-wavelength 
region around the Na I line to a transit light curve covering a wider wavelength region 
around the feature. For both transit observations, these two approaches
result in narrow-band spectra that are consistent with flat lines, and we did not 
detect narrow-band Na absorption. 
Scaling our narrow-band uncertainties to the bin size of \citet{nik14}, we expect that their 
3.3$\sigma$ Na I narrow-band absorption detection would translate into a 2.1$\sigma$ detection in our 
R150 observation, which is marginal, and therefore we did not consider our nondetection to be in 
disagreement with the HST study.

Another aspect of our observations is that they are seeing limited 
($\sim1.5$\arcsec), corresponding to a resolution element 
(line-spread function width) of $\sim3.5$\,nm near 
a wavelength of 600\,nm. This is similar to the bin size used 
by \citet{nik14} and also \citet[][3\,nm bin centered on 589\,nm]{nik16}, for instance, who observed the core
of the Na line on the warm Saturn-mass exoplanet WASP-39b 
using ground-based transit spectroscopy 
with the FOcal Reducer/low dispersion Spectrograph 2 (FORS2) 
instrument on the Very Large Telescope (VLT).

Both observations in principle also cover the K I resonant 
doublet at 770\,nm. This wavelength is strongly affected by telluric
O$_2$ absorption, however, and our systematic corrections are not reliable. We therefore excluded these data from consideration. 

\subsection{Stellar variability}
\label{sec:starvar}
Star spots and surface inhomogeneities have been known to affect the apparent
transit depth at a given wavelength, potentially affecting the transit 
spectrum of a transiting exoplanet. Star spots occulted by the planet during transit
have been investigated in detail \citep[e.g.,][]{des11a, des11b, nut11, san11}. 
We see no evidence of star spot occultations during transit in either of our white-light curves.
More recently, \citet{mcc14} and \citet{rac17}, for example,
have shown that unocculted star spots can
mimic a Rayleigh-like slope in the transit spectra of exoplanets due to 
what the latter authors called the transit light-source effect (TLSE). 
Faculae have been demonstrated to potentially affect the transit 
depths near strong stellar absorption lines such as H$\alpha$, 
Ca~II~K, and Na~I~D \citep[e.g.,][]{cau18, rac19}.
\citet{esp19} analyzed six transit spectroscopy observations of the hot Jupiter 
WASP-19b and characterized the surface of the host star. They detected 
two spot-crossing events, including a planetary occultation of a bright spot.  

However, HAT-P-1 and its companion (which is our reference star) 
have not been reported to show signs of activity so far. 
\citet{bak07} reported low atmospheric variability for both stars. 
Measurements of the Ca II H\&K activity index of HAT-P-1, $\log(R^{\prime}_{HK}),$
have yielded a value of $-4.984$, which indicates relatively low activity
\citep[][]{noy1984, knu10}. \citet{nik14} monitored 
HAT-P-1 for 223 days, between 2012 May 4 and 2012 December 13, using the 
RISE camera on the Liverpool Telescope.
This campaign covers our R150 observation on November 12, and \citet{nik14} 
placed an upper limit on the variability of the star of 0.5\%. 

\begin{figure*}[t]
  \centering
  \includegraphics[scale=0.70]{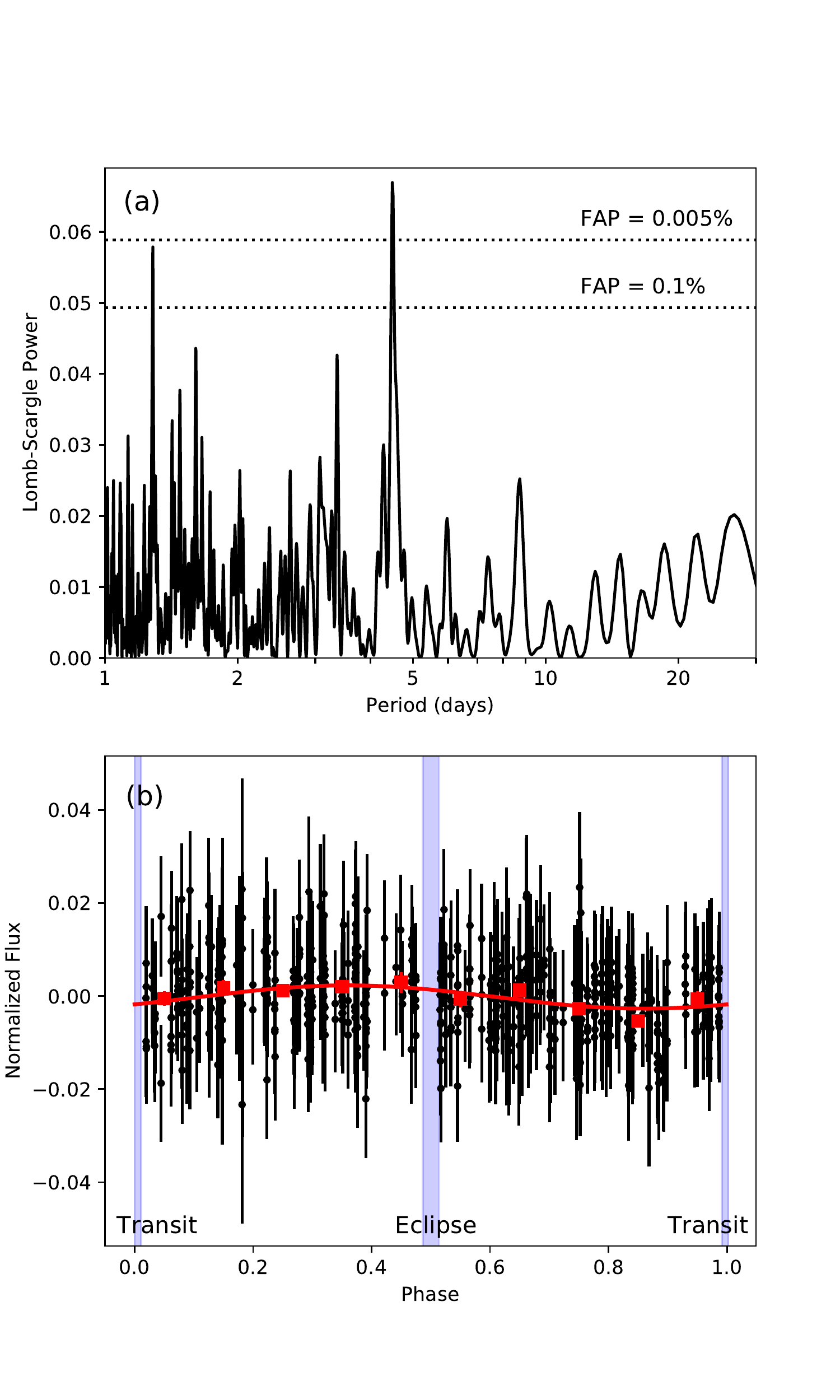}
  \caption{(a) Lomb-Scargle periodogram of our LCO B-band
  long-term photometric monitoring of HAT-P-1b (in-transit and in-eclipse measurements excluded). 
  The only interesting peak in the periodogram, near 4.4866\,days, with an FAP of $\sim4$\,ppm, corresponds to the 
  orbital period of the planet. The other significant peak, near 1.3 days, is likely 
  an alias of the LCO photometric cadence (typically one measurement every eight hours). 
  The rotation period of the star is expected 
  to be $\lesssim15$ days. (b) Our 
  stellar photometry folded to the period of
  the planet (4.465\,days) and fit with a sine curve. Red squares
  represent the photometry binned to $\sim11$ hours (0.1 of the orbital 
  period of the planet). 
  The variation (minimum to maximum) is about 0.5\%, which is smaller than the transit depth.  
  It varies slowly and therefore probably does not affect our transit 
  spectrum measurements. The timings of the transit 
  and the secondary eclipse are marked.}
  
  \label{fig:lomb_scargle}
\end{figure*}

We also examined the potential for variability and activity of the host star. 
We used our LCO photometric monitoring in the Johnson-Cousins/Bessell B-band, which should be 
particularly sensitive to magnetic star spots. 
Even though our photometric monitoring covers eight months in 2016 (Section~\ref{sec:lcodata}) and
does not cover our transit observations, any detection of variability in the B band 
would indicate that star spots and the TLSE effect need to be taken into account. 
We removed the photometric points taken during transit and secondary eclipse (20 and 37 points, respectively). We created a Lomb-Scargle periodogram using the remaining 617 data points in order to 
identify the stellar rotation period and the amplitude 
of variability, thus estimating the fraction of the stellar
surface covered by spots (Figure~\ref{fig:lomb_scargle}). 
However, we find no peak in the periodogram near the expected 
stellar rotation period \citep[$\lesssim15$\,days, based on 
rotational line broadening,][]{joh08}. This suggests
that the star had few if any stellar spots during the period it was monitored. 
We find no evidence for stellar activity in our photometric and spectroscopic data, nor in the literature. We therefore do not consider it necessary to correct our results 
for the transit light-source effect.

Our periodogram in Figure~\ref{fig:lomb_scargle} suggests a tentative low-amplitude periodic variation 
\citep[0.5\% in the B band, consistent with][]{nik14}
that coincides with the planetary orbital period. It is 
unlikely that this signal is an alias of the stellar rotation period because we do not detect its primary order. 
We speculate that this variability might
be an indication of a magnetic planet-star interaction, although
in this case, the variability period would be expected to be the 
synodic period of the star in the frame of the planet 
\citep[$\lesssim6.3$\,days, see, e.g.,][]{fis19}.

\section{Discussion}
\label{sec:mod}
We compare our observed transit spectrum to models 
computed using the Exo-Transmit code \citep{kem16} and to previous studies. The uncertainties of the spectrophotometry in the B600 observations exceed the
photon noise by a factor of up to four. 
This is likely an effect of the complex systematic effects that influence 
this data set, compounded by the high airmass toward the end of the 
transit observation. Consequently, the resulting B600 transit spectrum is consistent with both cloudy and clear models, and does not have the necessary precision to constrain the transit spectrum of HAT-P-1b.
Because the effect of the airmass on the data quality is especially strong
at wavelengths $\lesssim500$\,nm,  
it is difficult to place limits on the Rayleigh slope 
in the atmosphere of HAT-P-1b based on these data. 

\begin{figure*}
  \centering
  \includegraphics[scale=0.51]{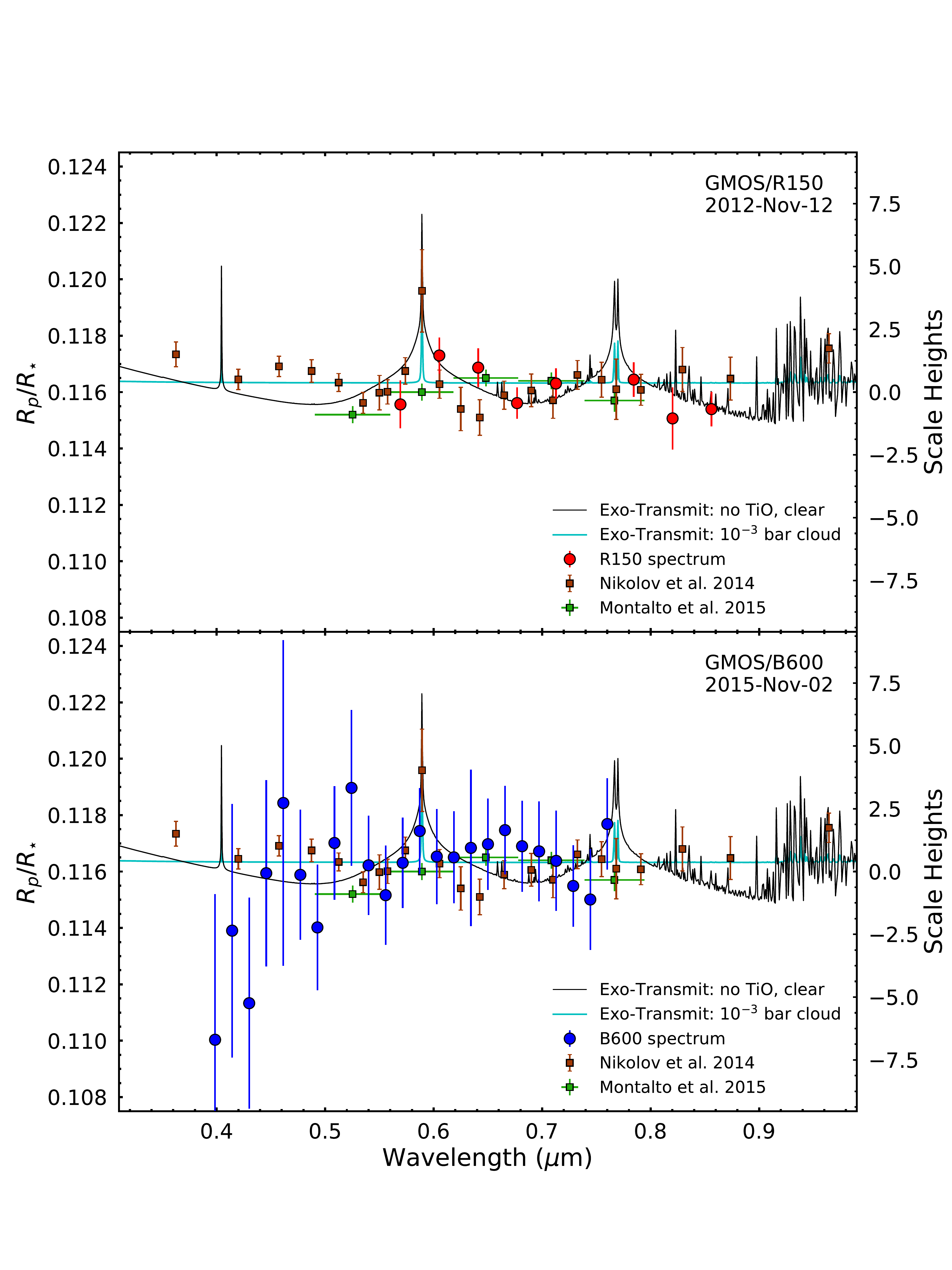}
  \caption{R150 (top) and B600 (bottom) transit spectra from this work, compared to the results 
  of \citet{nik14} and \citet{mon15}. The bluest wavelengths in the B600 spectrum are 
  strongly affected by high airmass, and this spectrum as a whole has little constraining
  power on atmospheric models. The R150 measurements near 750\,nm and $\gtrsim850$\,nm 
  could be affected by additional correlated noise effects of strong and variable telluric absorption
  by O$_2$ and H$_2$O, respectively. We therefore exclude these transit measurements from 
  our analysis and this figure. We compare the resulting transit spectrum to 
  an exo-transmit model with clear atmosphere and no TiO/VO (black lines), which, based on our observations, 
  is marginally favored over cloudy models, which are essentially flat lines (cyan lines). 
  The model that does not include 
  gas condensation and contains TiO/VO (not shown for clarity) is marginally disfavored by the R150 data.
  The R150 measurements have comparable uncertainties to the HST/STIS measurements (brown squares)
  for wavelength bin sizes of of tens of nanometers.}
  \label{fig:compare_all}
\end{figure*}

The R150 observation is less affected by 
systematic effects and typical uncertainties are only $\sim50$\%
above photon noise. The transit depths measured in R150 are comparable 
in terms of precision to the measurements made with HST/STIS, and the two 
observations are consistent with each other (Figure~\ref{fig:compare_all}). 

We compared the R150 reduced-$\chi^2$
values for several exo-transmit models (Table~\ref{tab:models}). All 
models we considered have isothermal atmospheres at 1500\,K, which is
close to the equilibrium temperature of the planet. Our spectrum excludes 
 strong TiO and VO features and shows preference to models with 
cloud decks deep in the atmosphere, or completely clear atmospheres. This is 
consistent with the HST/STIS results in \citet{nik14} and \citet{sin16}, 
and also with ground-based broadband transit measurements with
TNG/DOLORES \citep{mon15}. Atmospheric spectra without clear 
TiO and VO absorption features are common in hot Jupiters with 
equilibrium temperatures similar to that of HAT-P-1b
\citep[e.g.,][]{des08, sin16}.

\begin{table}[h]
\caption{R150 model results}
\begin{center}
\begin{threeparttable}
\begin{tabular}{ll}
\toprule
Model parameters & $\chi^2_{reduced}$ \\
\\
\hline\\[-1.5ex]
TiO, clear, no condensation &  2.9 \\
no TiO/VO, clear & 0.94  \\
no TiO/VO, cloud at 0.01\,mbar & 1.4  \\
no TiO/VO, cloud at 0.1\,mbar  & 1.4  \\
no TiO/VO, cloud at 1\,mbar    & 1.4  \\
no TiO/VO, cloud at 10\,mbar   & 1.2  \\
no TiO/VO, cloud at 100\,mbar  & 0.95 \\
\bottomrule
\end{tabular}
\begin{tablenotes}
\end{tablenotes}
\end{threeparttable}
\end{center}
\label{tab:models}
\end{table}

We tested whether our observations can be used to probe narrow-band features like 
the Na I resonance doublet at 589\,nm. Unlike \citet{nik14}, we did not detect 
the core of the Na absorption line, but our uncertainty estimates on the narrow-band 
transit depth uncertainties are $\sim60\%$ larger than those derived from the HST 
observations, which might explain this difference. 
We did not observe the broad line wings either, but this is consistent with 
the results of \citet{nik14}, who also did not detect them.

\section{Conclusion}
We presented ground-based observations of the 
visible low-resolution transit spectrum of the hot Jupiter HAT-P-1b. 
We showed that transit spectroscopy observations 
from the ground can be comparable to HST/STIS in precision and 
in information content, as is the case of our R150 observation. Based on 
our R150 observations between 550 and 850\,nm, we are able to 
reject models with a high abundance of TiO and VO. 
This is consistent with the findings
of previous space-based studies with HST/STIS. 
The B600 transit data are affected by strong systematic effects
and result in a highly uncertain transit spectrum. 

Our long-term photometric B-band monitoring of the host star, HAT-P-1, 
suggests that its activity is very low and is unlikely to affect
our spectrophotometry. However, our Lomb-Scargle 
diagram results in an unexpected peak at the planetary orbital period. 
Because of this, we speculate that planet-star interactions might play a role in this system.

\begin{acknowledgements}
We thank the referee for a careful and balanced review. 
We thank Jacob Arcangeli, Claire Baxter, and Gabriela Muro-Arena 
for useful discussions, and Nikolay Nikolov 
for graciously sharing with us the HST/STIS white light curves. 
This work is based on observations obtained at the Gemini Observatory
(acquired through the Gemini Observatory
Archive and Gemini Science Archive), which is operated
by the Association of Universities for Research in
Astronomy, Inc. (AURA), under a cooperative agreement
with the NSF on behalf of the Gemini partnership: the
National Science Foundation (United States), the
National Research Council (Canada), CONICYT (Chile),
Ministerio de Ciencia, Tecnolog\'{i}a e Innovaci\'{o}n
Productiva (Argentina), and Minist\'{e}rio da Ci\^{e}ncia, Tecnologia
e Inova\c{c}\~{a}o (Brazil).
Based in part on Gemini observations obtained from
the National Optical Astronomy Observatory (NOAO)
Prop. ID: 2012B-0398; PI: J-.M D\'{e}sert.
This work makes use of observations from the LCOGT network.
J.M.D acknowledges support by the Amsterdam Academic Alliance (AAA) Program.
The research leading to these results has received funding from the European Research
Council (ERC) under the European Union’s
Horizon 2020 research and innovation programme (grant agreement no. 679633;
Exo-Atmos). This material is based upon work supported by the
National Science Foundation (NSF) under Grant No. AST-1413663, and
supported by the NWO TOP Grant Module 2 (Project Number 614.001.601).
This research has made use of NASA’s Astrophysics
Data System. This research made use of Astropy, a community-developed core
Python package for Astronomy (Astropy Collaboration, 2018).
Other software used:
diff\_atm\_refr.pro
(http://www.eso.org/gen-fac/pubs/astclim/lasilla/diffrefr.html);
ATLAS \citep[][available at http://kurucz.harvard.edu]{kur93};
MPFIT \citep{mar09};
Time Utilities \citep{eas10}.
\end{acknowledgements}

%-------------------------------------------------------------------
\bibliographystyle{aa} % style aa.bst

\bibliography{hat-1_pre_v1.bib}

\end{document}